\documentstyle[12pt,epsfig]{article}
\begin{document}

\title{ Efficiency of the Incomplete Enumeration algorithm for
Monte-Carlo simulation of linear and branched polymers.\\
\vspace{0.2cm} Sumedha\thanks{sumedha@theory.tifr.res.in}\\ and\\
Deepak Dhar\thanks{ddhar@theory.tifr.res.in}\\
\vspace{0.2cm} Department Of Theoretical Physics\\ Tata Institute Of
Fundamental Research\\ Homi Bhabha Road, Mumbai 400005\\ India} \date{}

\maketitle

\begin{abstract}
          We study the efficiency of the incomplete enumeration
algorithm for linear and branched polymers. There is a qualitative
difference in the efficiency in these two cases.  The average time to
generate an independent sample of configuration of polymer with $n$
monomers varies as $n^2$ for linear polymers for large $n$, but as
$\mbox{exp}(c n^{\alpha})$ for branched (undirected and directed)
polymers, where $0<\alpha<1$. On the binary tree, our numerical
studies  for $n$ of order $10^4$ gives $\alpha = 0.333 \pm 0.005$. We
argue  that $\alpha=1/3$ exactly in this case.
\end{abstract}

Monte-Carlo(MC) simulations are a very important tool for studying
polymers, as exact results are hard to come by, and are available only
for the simplest models. Broadly speaking, MC algorithms fall  in two
classes \cite{sokal}: the Metropolis type and the genetic type.  The
Metropolis type algorithms generate a time sequence of  configurations
of the polymer using a Markovian evolution. The transition
probabilities from one configuration to the next are so chosen that
the  time average of properties of the system are equal to that from
the  desired distribution. These may use local moves as in Rouse
dynamics  \cite{rouse}, bi-local moves as in the reptation algorithm
\cite{reptation}  or nonlocal moves as in the pivot \cite{pivot}  and
cut-and-paste \cite{cutpaste} algorithms. There is inevitably some
correlation between different configurations generated in an
evolution.  These algorithms become inefficient if the correlation
time becomes  very large, eg. when simulating polymers in a random
medium.

In the genetic algorithms, one randomly generates a small random
number of configurations in each run. The probability that a given
configuration is obtained in a run is proportional to the desired
distribution. One repeats the process for many runs to get a large
sample. Examples of this type are the enrichment \cite{wall} and  the
pruned-enriched Rosenbluth method(PERM)-like \cite{grassberger}
algorithms.

While there have been many studies of linear polymers  using various
Monte-Carlo techniques like pivot \cite{pivot,saw,kennedy},  PERM
\cite{grassberger}, Berreti-Sokal algorithms \cite{berreti},  branched
polymers have been less studied. Algorithms used for simulating linear
polymers can often be adapted for branched polymers, but they are
usually  found to be less efficient. For example, in the pivot
algorithm,  the acceptance probability of the transformed
configuration is found to  be much less for branched polymers than for
linear polymers \cite{rensburg}.  The algorithm does not perform well
for branched polymers adsorbed on  a surface\cite{you}. The PERM
algorithm also seems to work  less well for branched polymers than for
linear polymers \cite{grassberger1}.  Incomplete enumeration(IE) is an
algorithm belonging to the genetic class of  algorithms. It has been
used for simulating linear polymers\cite{redner},  and for branched
polymers \cite{dhar,lam}.

A better understanding of the efficiency of Monte Carlo algorithms for
generating branched polymers seems desirable. We will study IE for
linear and branched polymers in this paper. We choose the  average
computer time $T_n$ needed to generate one statistically  independent
sample of desired size $n$ as a reasonable measure of  efficiency of
the algorithm. The dependence of $T_n$ on $n$ is very  different for
IE for linear and branched polymers. We find for the linear  polymers
$T_n \sim k n^2$, but for branched polymers  $T_n \sim \mbox{exp}(k
n^{\alpha})$, $0<\alpha<1$. We also discuss an  improvement of IE
which we call improved incomplete enumeration (IIE), in this paper. We
find that the improvement  does not change the asymptotic  dependence
of $T_n$ on $n$ in general. IIE works better than IE but  the
difference is only in the coefficient $k$.

The plan of the paper is as follows. We describe the IE algorithm in
Section 1. In Section 2 we discuss the efficiency criterion for MC
algorithms in general, and for IE in particular. In Section 3 we study
the efficiency  of IE analytically for some simple cases where  the
genealogical tree has a simple recursive structure.  We also study IE
for self avoiding walks(SAW) in this section. In all  cases we find
that $T_n \sim n^2$. In Section 4 we propose an improved  version of
the IE algorithm, IIE. For simple random walks $T_n = n$  for the IIE
algorithm as compared to $T_n \sim n^2$ for IE. For SAWs, IIE is
significantly more efficient and  becomes better in higher dimensions,
but asymptotic efficiency remains  the same and $T_n \sim a_d n^2$ in
all dimensions, though the coefficient $a_d$ decreases with increasing
dimension. In Section 5 we study IE for branched polymers or lattice
animals on a binary tree. We give heuristic  arguments and numerical
evidence to show that $T_n \sim \mbox{exp}(k n^{1/3})$ for large $n$
for branched polymers on a binary tree. We also study IE  and IIE
numerically for undirected and directed branched polymers on a square
lattice in this section. We find that in both cases  $T_n \sim
\mbox{exp}(k n^{\alpha})$, $0<\alpha<1$. We summarise our  results in
Section 6.

\section{The Incomplete Enumeration Algorithm}

Self-avoiding walk and lattice animals(LA) are simple lattice models
of linear and branched polymers in dilute solutions. In order to
study the thermodynamic properties of these polymers, one has to
average over all allowed configurations of the polymer of a given
number of monomers. The averages are defined with all configurations
considered to be equally likely. Since the total number of possible
configurations grow exponentially fast with size of the polymer,
brute-force exact calculation is possible only for small polymers.
Monte-Carlo methods allows us to study much larger sizes by obtaining
a representative sample of the set of configurations and estimate the
ensemble averages from the sample average.

The IE algorithm is a simple modification of exact enumeration
algorithm for generating polymers. A good exact enumeration algorithm
generates all possible configurations exactly once \cite{martin}. This
is ensured by defining a rule which, given an $n$-site configuration
of a polymer, identifies uniquely one of these sites as the `last
added site'. Removing this site must result in an allowed polymer
configuration of $(n-1)$ sites. The $(n-1)$-site polymer is called the
parent of the $n$-site configuration. We start by imagining that we
have  arranged all configurations in a genealogical tree, whose nodes
are  the different configurations of the polymer, such that all
polymer  configurations of $n$ sites are at level $n$ and are
connected to their  parent at level $(n-1)$. Clearly, the tree depends
on the rule used to define parenthood. For example, Fig. \ref{extree}
shows a genealogical tree for directed lattice animals on a square
lattice for $n \leq 4$, using one such choice (see Appendix for
details). In the actual implementation  of the algorithm, the whole
genealogical tree is not constructed first, and  tree is constructed
and the pruning is decided as we proceed in a depth first search.

As the number of configurations of polymer of size $n$ increases
exponentially with $n$, the time required to construct the
genealogical tree up-to level $n$ in the exact enumeration algorithm
increases exponentially with $n$. The basic idea of the IE algorithm
is to decrease this time by randomly pruning the genealogical tree.

In IE we choose a set of $(n-1)$ real numbers $p_i$ ($0 < p_i \leq
 1$), for $i=1$ to $(n-1)$. Any bond in the genealogical tree
 connecting level $r$ to level $(r+1)$ is removed with probability
 $(1-p_r)$ independent of the other bonds. If a configuration gets
 disconnected from the root node, automatically all its descendants
 are also removed. We make a  depth first search of the pruned
 genealogical tree up-to depth $n$ to  determine the different
 configurations that remain at level $n$. We run  the algorithm
 several times to generate a large sample. The probability  of
 enumeration of a particular $r$ site configuration in a given run is
\begin{equation}
{\Xi}_r = \prod_{i=1}^{i=r-1} p_i  \label{Pin}
\end{equation}

This is same for all configurations of size $r$. This ensures that the
sample of configurations obtained is unbiased. As a configuration can
occur at most once in a single Monte-Carlo run, IE samples the
population without replacement.

The different runs are mutually uncorrelated. However, the number of
configurations produced within one run varies from run to run, and
different configurations produced in the same run are
correlated. Also, the fraction of runs in which one generates at least
one configuration of size $n$ goes down with increasing $n$.

In case of SAWs which model linear polymers, there is a natural
labelling scheme in which one just labels the first point of walk by
$1$, the second by $2$ and so on. In case of branched polymers there
are several different choices of labelling possible corresponding  to
different possible rules of removing a site from a $n$-site cluster
to generate a $(n-1)$-site connected cluster. We have used the
Martin's labelling scheme \cite{martin} for our cluster counting
algorithms. A brief description of this can be found in the Appendix.

\section{Efficiency}

In general, in Monte-Carlo methods, the time needed to estimate an
ensemble average $\mu = \langle {O}\rangle$ of some observable $O$
over all clusters of size $n$ averaged over $N$ independent samples
would give estimate as $\mu^* = \mu \pm \sigma/\sqrt{N}$, where
$\sigma^2$ is the  variance of $O$. If correlations are present,  the
average time required to estimate $\mu$  within the fractional  error
$\epsilon$ varies as $(\sigma/{\epsilon} \mu)^{2} \tau$,  where $\tau$
is a measure of correlations in the data. For Metropolis  evolution,
$\tau$ is the auto-correlation time of the observable $O$.  In the
case of IE, the efficiency depends on the average time taken  by the
Monte Carlo algorithm to generate a single run and the degree of
correlations present in the different samples produced in the same run.

It is difficult to determine the latter exactly for IE. It  depends
also on the quantity we want to average. Consider a set of
configurations generated by $N$ independent runs of the IE
algorithm. Let the probability that a single run produces at least one
sample be $P(n)$, and the average number of configurations produced
per run be $a$. Then for large $N$, we will generate approximately
$Na$ configurations, which will be made of approximately $P(n) N$
mutually uncorrelated groups. Thus the average size of a correlated
group is $a/P(n)$. It seems reasonable to measure the efficiency of
the  algorithm in terms of the average CPU time required to produce
one  independent group of configurations. This overestimates
correlations  as this treats all samples produced within one run as
fully correlated
\footnote{For example, the mean is $2749.2$ and the standard deviation 
$\sigma$ is $1349.2$ for the radius of gyration of animals of size $50$ 
for the full population. The average number of samples produced per 
successful run of MC simulation was $27.5$. If we calculate the average 
radius of gyration of 100 samples of $10^4$ consecutive runs, the 
standard deviation $\sigma^{'}$, of the mean calculated is $144.2$. 
This would have been $\sigma/100$ if they were uncorrelated. Similarly, 
for SAW of size $50$ on a square lattice the average number of samples 
produced per successful run is $5.3$ and $\sigma = 60.44$, and 
$\sigma^{'} = 1.4$ for $100$ samples of $10^4$ consecutive runs.}.

Other definitions of  efficiency are possible, and may be
advantageous in specific contexts. For example, one may be interested
in  some asymptotic properties of the polymer problem, like the
branching number  $\lambda$, or the critical exponent $\theta$. In
this case, the value of $n$  is not decided beforehand, and the
desired estimate is obtained by suitable extrapolation of data for
different $n$. We can study average  number of descendants $<X_n>
\approx \lambda(1-\theta/n)$ to estimate  $\lambda$ and
$\theta$. Analysis of errors in such quantities is more  complicated,
and will not be discussed here.

Let $T_n$ be the average CPU time required to obtain one run which
generates at least one configuration of size $n$. If $\tau_n$ is the
average CPU time for one Monte-Carlo run, then we have

\begin{equation}
T_n  =  \frac{\tau_n}{P(n)}
\end{equation}

The average CPU time required for one run is estimated easily in terms
of the time taken to add or delete a configuration on the genealogical
tree. We define this to be one unit of CPU time.

The total CPU time for one MC run is proportional to the number of
nodes  in the pruned genealogical tree. Let $X_j$ denote the random
number of  $j$ site configurations generated in a single run. The time
to visit sites  of the randomly pruned tree up-to depth $n$ is
$\sum_{j=1}^{n} X_j$.  The CPU time in a run is then proportional to
the number of nodes  in the pruned tree. The average CPU time per run
$\tau_n$, would be  equal to the sum of average values
$\langle{X_j}\rangle$, averaged over  all runs.

\begin{equation}
\tau_n \propto \sum_{j=1}^{n} <X_j>
\end{equation} 

For linear and branched polymers, the total number of configurations
$A_n$ of a given size $n$ is known to vary as

\begin{equation}
A_n \sim A {\lambda}^n n^{-\theta}
\end{equation}
for large $n$. Here $A$ is a constant, $\lambda$ is called the growth
constant and $\theta$ is a critical exponent.  Since each
configuration  with $n$ sites has a probability $\Xi_n$
(Eq. (\ref{Pin})) of being  generated, and there are $A_n$ total
number of configurations,  $\langle{X_n}\rangle = {\Xi}_nA_n$, giving

\begin{equation}
\tau_n = \sum_{j=1}^{n} A_j \Xi_j
\end{equation}

Since $\langle{X_n}\rangle$ can be directly estimated in IE, we get a
way  to estimate the number of configurations $\langle{X_n}\rangle$ by
simulations. This can be used to estimate the $\lambda$ and $\theta$.

A study of the efficiency of the algorithm is complicated as $P(n)$
depends on the structure of the genealogical tree, and is difficult to
determine theoretically.

An upper bound on working of these algorithm is the time for exact
enumeration of all the samples, which is exponential in $n$. Consider
the case in which $p_i=p$ for all $i$. So long as  $p \lambda > 1$,
$\langle{X_n}\rangle$ will grow exponentially with $n$. As $P(n)\leq
1$, this implies that $T_n$ increases exponentially with $n$ if $p
\lambda >1$. Also, if $p \lambda <1$, then $P(n)$ varies as $(p
\lambda)^n$ to leading order, but $\tau_n$ remains finite ($\tau_n
\geq \tau_1$) \cite{harris}. Thus again $T_n$ increases exponentially
with $n$.  These two considerations together imply that a good choice
of $p$ is that it  should be approximately equal to
$1/\lambda$. However, finding the  optimal choice of $\{p_i\}$ for a
given problem is non trivial.  We investigate this in the next section
for some illustrative cases.

\section{Optimising the IE algorithm}

\subsection{Systems with Uniform genealogical tree}

The simplest of enumeration problems is the enumeration on a uniform
genealogical tree. For example, random walks which are models for
linear polymers without self exclusion correspond to a uniform
genealogical tree of branching number $\lambda$. The number  of nodes
at level $n$ is $\lambda^{n-1}$.

Consider a uniform genealogical tree with two descendants per node. In
this case number of nodes at level $n$ would be $2^{n-1}$. For the
choice of  $\{p_i\}$, the probability of connection of root with level
$r$, denoted  by $P(r)$ follows a simple recursion relation

\begin{equation}
P(r+1) = 2p_r P(r) -p_r^2 P^2(r) \label{urec}
\end{equation}
with $P(1)=1$. The average CPU time per run $\tau_n$ is given by

\begin{equation}
\tau_n = 1+\sum_{i=2}^{n} 2^{i-1} \Xi_i
\end{equation}

First we try to find out as to what choice of $p_i's$ minimises $T_n$
for small $n$.

For small sizes one can try systematic optimisation. Let us choose
$n=2$.  Then on the binary tree, $P(2)=2p_1-{p_1}^2$ and $\tau_2=1+2
p_1$. This  gives

\begin{equation}
T_2 =  \frac{2p_1+1}{2p_1-{p_1}^2}
\end{equation}

Minimising with respect to $p_1$, we get the minimum value of $T_2$ to
be $(3+\sqrt{5})/2 \approx 2.618$ for $p_1 =(\sqrt{5}-1)/2 \approx
0.618$.

Similarly, the time ($T_3$) of IE for reaching level $3$ from level
$1$, is given by

\begin{equation}
T_3 =\frac{1+2p_1+4p_1p_2}{2p_1 (2p_2-p_2^2)-p_1^2 (2p_2-p_2^2)^2}
\end{equation}

It is easy to check that $T_3$ in this case takes its minimum value
for $p_1=0.534$ and $p_2=0.618$. Similarly for $n=4$, the minimum
occurs  at $p_1 = 0.516$, $p_2=0.534$  and $p_3 =0.618$. For large
$n$,  the best choice of $p_i$ tends to $1/2$. By optimising till
$n=30$, we  find that the best choice of $p_i$ is quite well described
by the  approximate formula $p_i \approx \frac{1}{2} (1+0.5/(n-i)^2)$.

For large $r$, if $p_r \rightarrow p^*$, Eq. (\ref{urec}) can be
approximated by $P(r+1) = 2 p^* P(r) - {p^*}^2 P(r)^2$. For $2p^*<1$,
we get $P(r) \rightarrow (2p^*)^r$ decreases exponentially with
$r$. For $(2 p^*)>1$, it leads to $P^*({\infty}) \sim (2p^*-1)$.

We have already argued that $p_i$ should be close to $1/\lambda$, else
the algorithm is inefficient, $T_n$ varies as
$\mbox{exp}(n)$. Consider now the case where
$p_i=\frac{1}{\lambda}(1+\alpha/i^{m})$, where $\alpha$ and $m$ are
parameters that we can vary to find the optimal values. In this case,
$\langle{X_n}\rangle =  \prod_i (1+\alpha/i^{m})$, and $P(n)$ is
approximately given by

\begin{equation}
\frac{\partial P(n)}{\partial n} = \frac{\alpha}{n^m} P(n) -
\frac{1}{\lambda^2} P^2(n)   \label{udiff}
\end{equation}

Then , if $m>1$, we see that $\langle{X_n}\rangle$ tends to a constant
for large $n$, and $\tau_n$ is proportional to $n$.  Also, $P(n)$
varies as $1/n$, and we have $T_n \sim n^2$.

If $m=1$, and $-1<\alpha<1$, then $\langle{X_n}\rangle$ varies as
$n^{\alpha}$, and hence $\tau_n \sim n^{\alpha+1}$. Also,
Eq. (\ref{udiff}) gives $P(n) \sim A (1-\alpha)
n^{\alpha-1}$. Interestingly, in the $T_n$, these powers cancel and we
get $T_n = \tau_n/P(n) \sim C_{\alpha}n^2$. We find that $C_{\alpha}
\sim 1/(1-\alpha)$, hence the best choice of $\alpha$ is $\alpha=0$.

If $m<1$, then $\langle{X_n}\rangle$ varies as $\mbox{exp}(n^{1-m})$,
and $P(n)$ varies as $n^{-m}$, and hence $T_n$ varies as
$\mbox{exp}(n^{1-m})$ to leading order, thus in this case $m<1$  leads
to a suboptimal performance of the algorithm.

On a binary tree for $p_i=\frac{1}{2}$, we get $T_n=n^2/4$. From
systematic optimisation we saw that there exist a nontrivial optimal
value for each $p_i$ which depends on the depth of the genealogical
tree to be reached. This value for uniform binary tree was $p_i
\approx \frac{1}{2} (1+0.5/(n-i)^2)$. But even with this choice for
large $n$ we get $T_n \approx n^2/4$. This result is generalised
straight forwardly to $k$-node uniform tree. For the choice,  $p_i =
1/k$ $\forall i$, we get  $T_n = \frac{(k-1) n^2}{2k}$.
                                                            
\subsection{Systems with recursively defined genealogical tree}

It is necessary to check how non-uniformity of trees can change the
 above conclusions. The simplest of non-uniform trees are the
 recursively  defined trees. The number of branches from a given node
 still follow  a definite pattern which repeats and depends on the
 coordination number  of the parent node. We consider some examples

\begin{figure}
\begin{center}
\epsfig{figure=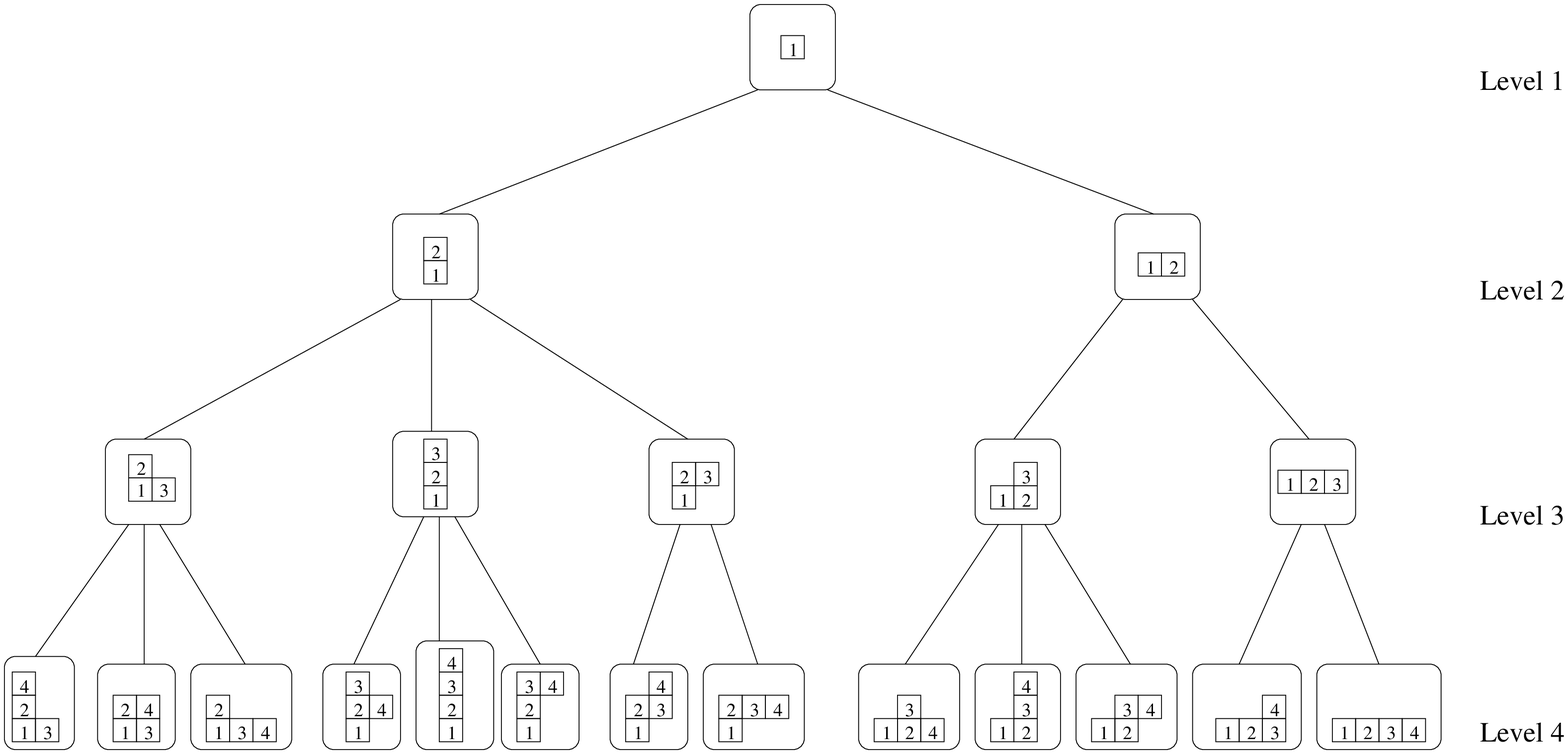,width=18cm,angle=90}
\end{center}
\caption{An example of a genealogical tree. The numbers labelling the
sites indicate the order in which they are added ($1$ represents the
root site). The tree shown is for directed lattice site animals on a
square lattice.}   \label{extree}
\end{figure}

A node with $k$ descendants will be called a $k$-node. Consider a tree
specified by the rule that the descendants of a $2$-node are  a
$2$-node and a $3$-node, and the descendants of a $3$-node are  one
$2$-node and two $3$-nodes. We specify such a tree by the notation
$(23,233)$ tree (Fig.\ref{extree}). If $B_2(n)$ and $B_3(n)$ are
respectively the number of nodes at level $n-1$ which have $2$ and $3$
descendants respectively, then

\begin{eqnarray}
B_2(n) = B_2(n-1)+B_3(n-1)\\ B_3(n) = B_2(n-1)+2B_3(n-1)
\end{eqnarray}

From these linear recursion equations it is easy to see that $B_2(n)$,
$B_3(n)$ and also the total number of nodes at depth $n$, $A_n$, all
grow as $(\lambda)^n$ for large $n$, where $\lambda = (3+\sqrt{5})/2)$.

We now look at the efficiency of IE on this tree. Take all $p_i = p$.
We define $P_2(r)$ and $P_3(r)$ as the probabilities that a 2-node and
a $3$-node respectively are connected to at least one node $r$ levels
\ below. Clearly they have the following recursions
\begin{eqnarray}
1-P_2(r+1) = (1-pP_2(r))(1-pP_3(r))\\ 1-P_3(r+1) =
(1-pP_2(r))(1-pP_3(r))^2
\end{eqnarray}
with $P_2(1)=P_3(1)=1$.

For large $r$, near the fixed point we get $P_2(r) \approx
\frac{p}{1-p} P_3(r)$. Substituting  in the second equation, we find
the linear term vanishes for $p=1/\lambda$ and the difference equation
can be approximated by $\partial P_2/\partial r \sim -P_2^2$, which
implies that $P_2(n)$ and $P_3(n)$ decay as $1/n$ for large $n$. We
get $P_2(n) \approx \frac{\lambda^2}{(1+\lambda)n} $. The total CPU
time at $p=1/\lambda$ is $\frac{(5+\sqrt{5}) n}{10}$. It gives the
upper bound on time per independent run to be
$\frac{(\lambda-2)(1+\lambda)}{(3 \lambda-2)}n^2 \approx 0.382n^2$.

We can similarly analyse the other recursively defined trees. Consider
for example, the tree given by the rule $(23,223)$. We find that
growth constant $\lambda$ is $2.4142$ and for $p_i=1/\lambda$ for IE
this gives $T_n \approx 0.396n^2$. On a $(33,233)$ with growth
constant  2.732 for $p_i=1/\lambda$ for IE this gives
$T_n=\frac{(\lambda+4)}{4(\lambda+2)}n^2 \approx 0.35n^2$. It is easy
to convince oneself that for all recursively defined trees we get $T_n
\sim n^2$.

\begin{figure}
\begin{center}
\epsfig{figure=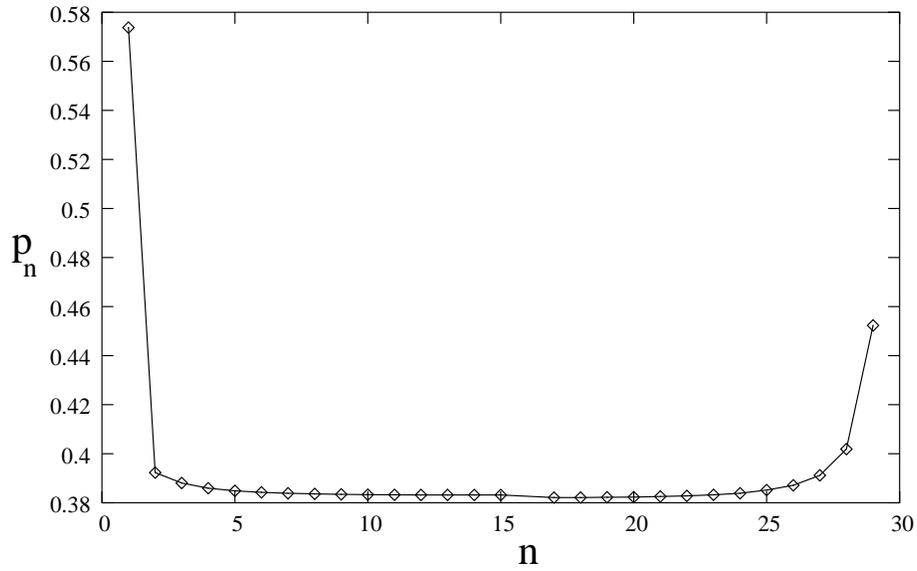,width=12cm}
\end{center}
\caption{Plot of optimum values of $p_i$ on a $(23,233)$ tree of depth
$30$}    \label{ran32}
\end{figure}

It is instructive to see the results of systematic optimisation over
$\{p_i\}$ in case of non uniform trees. Similar analysis for (23,233)
tree (Fig. \ref{extree}) between level 1 and 2 gives $p_1
=0.618$. Similarly optimising $T_3$ between level 1 and 3 gives
$p_1=0.562$ and $p_2=0.484$. An optimisation between level $1$ and $4$
gives the best values of $p_i's$ to be $p_1 = 0.562$, $p_2=0.42$ and
$p_3 = 0.467$. We see that the optimal value of $p_i$ in this case
depends on $n$. By optimising till $n=30$, where $n$ is the depth of
the genealogical tree, we find that for tree levels away from root and
bottom, optimal value of $p_i$ approaches $1/\lambda$ with increasing
$i$ and the asymptotic behaviour of algorithm remains the same as long
as we choose $p_i \approx 1/\lambda$. The optimal $p_i$ values as a
function of $i$ are plotted in Fig.\ref{ran32}. The optimising value
of $p_i$ are a bit higher than $1/\lambda$ near the two ends $i=1$ and
$i=n$. This extra optimisation does not change the $T_n \sim K n^2$
dependence, and infact does not change the asymptotic value of $K$
either.

The incomplete enumeration algorithm generates a bond percolation
process on the genealogical tree, where each link  is present
independently with a probability $p$. We define the  percolation
threshold $p_c$ on the tree to be such that for all  $p>p_c$, there is
a non zero probability that the starting node  belongs to an infinite
cluster. For $p<p_c$ the probability of  connection between root and
level $n$ usually goes down exponentially  in $n$. At $p_c$ it is
expected to decrease as a power law in $n$ and  for $p>p_c$ it takes a
finite value in the limit of  $n \rightarrow \infty$. The $p_c$ on a
tree is bounded from below  by $1/\lambda$ \cite{grimmett}. For the
genealogical trees which  we discussed, the $p_c$ was equal to
$1/\lambda$ and the optimal  behaviour of the algorithm was achieved
for $p_i \approx 1/\lambda = p_c$.

\subsection{Self avoiding Walks}

We now consider IE for SAW. For a SAW on a $d$ dimensional lattice,
 the number of configurations $A_n \sim \lambda^n n^{\gamma-1}$, where
 $\lambda$ is a lattice dependent constant and $\gamma$ depends only
 on the dimension.  The exponent $\gamma$ is known to be $1$ for
 $d>4$, and $\gamma = 43/32$ for $d=2$ \cite{madras}. The exact value
 of $\lambda$ is known for the hexagonal lattice \cite{nienhuis}, and
 fairly precise numerical  estimate, which matches well with root of a
 quartic equation with integer coefficients is known on the square
 lattice \cite{jensen1}.

\begin{figure}
\begin{center}
\epsfig{figure=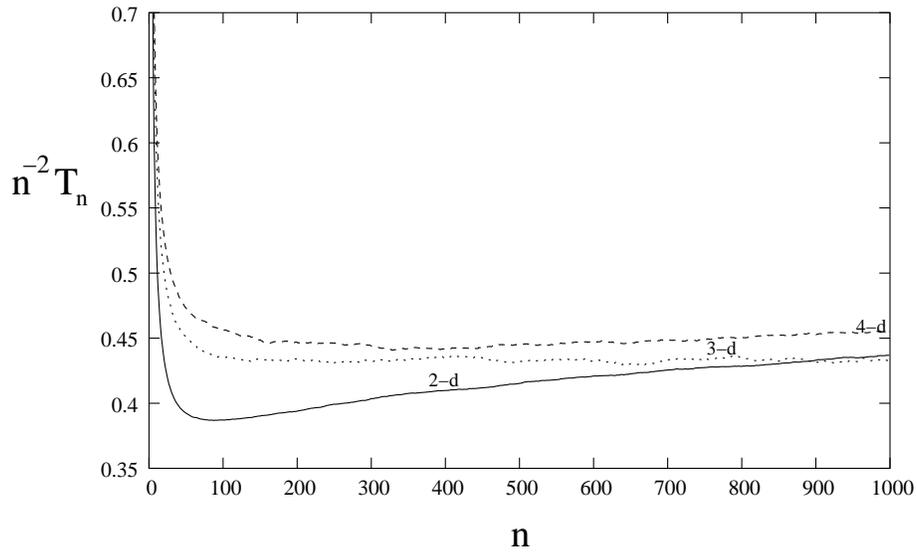,width=12cm}
\end{center}
\caption{$T_n/n^2$ of IE as a function of size $n$ for SAW on a
  $2$,$3$ and $4$ dimensional hyper cubic lattice. The lower most
  graph is for SAW on a square lattice and middle one in $3-d$ and the
  topmost is for $4-d$.}   \label{sawin}
\end{figure}

The genealogical tree for SAW is not uniform. For example, for rooted
 SAW(one end fixed at origin) on a square lattice, the number of
 different allowed choices of the $n^{th}$ step for $n>1$ varies  from
 0 to 3, depending on the walk. In this case it is difficult to
 determine the probabilities of connection up-to level $n$
 analytically but we have estimated $P(n)$ numerically by
 simulations. We choose  $p_i = \lambda^{-1}(1+1/i)^{1-\gamma}$, so
 that on the average we get order one configurations of size $n$ per
 run for large $n$. With this choice of $p_i$ our numerical
 simulations show that the probability of reaching level $n$ goes down
 as $1/n$ and hence whenever level $n$ is  reached, on an average
 $\sim n$ SAWs of size $n$ are generated. This also  implies that
 $p_c$ is indeed $1/\lambda$ on the SAW genealogical tree. We did
 $10^6$ Monte-Carlo simulations and generated walks up-to size
 $10,000$ on a square lattice. We have plotted $T_n$ in Fig
 \ref{sawin}. Our numerical fit suggests $T_n$ for IE to be $(0.42\pm
 0.01) n^2$.

In 3 dimensions $\lambda =4.6839 $ and $\gamma=1.16$ \cite{madras} and
nearly $90\%$ nodes have coordination number $5$. Hence the tree  is
more uniform than the $2d$ case and we get $T_n \approx 0.43 n^2$
(Fig. \ref{sawin}).

The genealogical tree becomes more and more uniform as we go to higher
 dimensions. In general on a $d$ dimensional hyper-cubic lattice the
 maximum branching possible is $2d-1$ and in the limit $d \rightarrow
 \infty$ the growth constant has an expansion \cite{madras}

\begin{equation}
\lambda =  2d-1-\frac{1}{2d}-\frac{3}{(2d)^2}-.....
\end{equation}

Hence the dominant branching is $2d-1$ and probability of a node
branching into $2d-1$ branches increases with dimension, and the lower
branching numbers occur with much smaller frequencies. The probability
of connection to level $n$ is hard to obtain analytically for any $d$.

In Fig. \ref{sawin} we have also shown a plot of efficiency of IE in
$3$ and $4$ dimensions for SAW. In few hours one can simulate $10^5$
Monte-Carlo runs for walks of size $1000$ on a Pentium-4 machine. We
get $T_n \sim n^2$ for $2$,$3$ and $4$ dimensions. This leads us to
conclude that the small non uniformity of the genealogical tree is
unimportant and $T_n$ varies as $n^2$ in all dimensions for SAW.

We note that for SAWs, other algorithms like pivot are known to be
more efficient. For pivot algorithm the correlation time for end to
end length varies as $n^x$ with $x<1$ in two dimensions
\cite{kennedy}. However, if we want to study some variable like
correlations in the directions of consecutive steps of the walk, the
correlation time will have to satisfy the inequality, $T_n \geq n$, as
one would need to update each step about  $O(1)$ times to affect the
nearest neighbour correlations.

\section{Improved Incomplete Enumeration (IIE)}

The main limitation of IE is attrition: the probability of generating
 $n$-site configurations in a given Monte-Carlo run goes down with
 $n$. One way to increase the probability of survival is to
 redistribute weight amongst the descendants in such a  way that while
 the probability that a particular node is selected remains same as
 before, the probability that at-least one of the descendants is
 chosen is increased. We call this `Improved Incomplete
 Enumeration(IIE)'.

Suppose in the implementation of IE as outlined in Section 1, we come
to a  node with degree $j$. Then in IE, each link is independently
deleted with a  probability $(1-p)$, and the probability that all
links are deleted is  $(1-p)^j$, which is non zero, even if the
expected number of descendants of  this node is $pj>1$. In IIE, the
links are not deleted independently. The  probability that any given
node is selected remain $p$, but the probability  that at least one
node is selected increases. This is implemented as follow:  If there
are $j$ descendants of a node and each link downward is present with
probability $p$, then  we choose Int$(pj)$ edges at random and give
them  weight one, and select one of the edge out of the remaining $j$
at random and  give it a weight one with probability frac($pj$) and
delete all the other  edges.

Hence we see that in IIE, though the average probability of selection
 of an edge remains $p$, but it enhances the probability of connection
 between two level of the genealogical tree and hence the probability
 of success in a given Monte Carlo run. For example, as will be
 discussed in the next section, on a regular tree with $p=1/\lambda$,
 the probability of connection up-to $n$ levels below in IIE is
 exactly one whereas it goes as $1/n$ in IE.

\subsection{Systems with recursively defined genealogical tree}

In IIE one redistributes the sum of probabilities of connection  from
a node to the next level. On a uniform binary tree  $y_i=2$ $\forall
i$ and with $p_i=1/2$, $y_i p_i=1$ and hence for $p_i=1/2$ with IIE
probability of reaching any level $n$ of the tree after $n$ steps is
exactly $1$ and exactly one configuration of any given size is
generated in the process and hence $T_n=n$. With $p_i =1/k$ this
result holds for any $k$ node uniform tree. Clearly $p_i =1/k$ is the
best choice in this case, as an absolute lower bound on time $T_n$ of
the algorithm is $n$.

If we use the improved algorithm for a (23,233) tree,
$\langle{X_n}\rangle$  and hence the average CPU time per run will
remain the same. We can also  determine the connection probabilities
$P_2(n)$ and $P_3(n)$. The coupled  difference equations for $P_2(r)$
and $P_3(r)$ have no cubic term. The  recursions are

\begin{eqnarray}
P_2(r+1) =  p(P_2(r)+P_3(r))\\ P_3(r+1) = p (P_2(r)+ 2
P_3(r))-\frac{3p-1}{3} (2 P_2(r) P_3(r) +P_3^2(r)) \label{32recursion}
\end{eqnarray}

		which at $p=1/\lambda=p_c$ gives $P_2(n)$ varying as
 $1/n$ for large $n$. The time per independent run comes out to be
 $\frac{\lambda (3-\lambda)}{3}\approx \frac{1}{3}$ times that in
 incomplete enumeration. That is, IIE is nearly three times more
 efficient than IE.

IIE certainly works better than IE. But, except for the uniform tree,
the difference between IE and IIE is only in the coefficient of
$n^2$. While performance of IIE improves as the genealogical tree
becomes more and more uniform, there is no qualitative difference  in
the efficiency of IE and IIE on a recursively defined non uniform
tree.

\subsection{IIE for SAW}

We studied IIE on a $d$ dimensional hyper-cubic lattice for $d =2$ to
10.

\begin{figure}
\begin{center}
\epsfig{figure=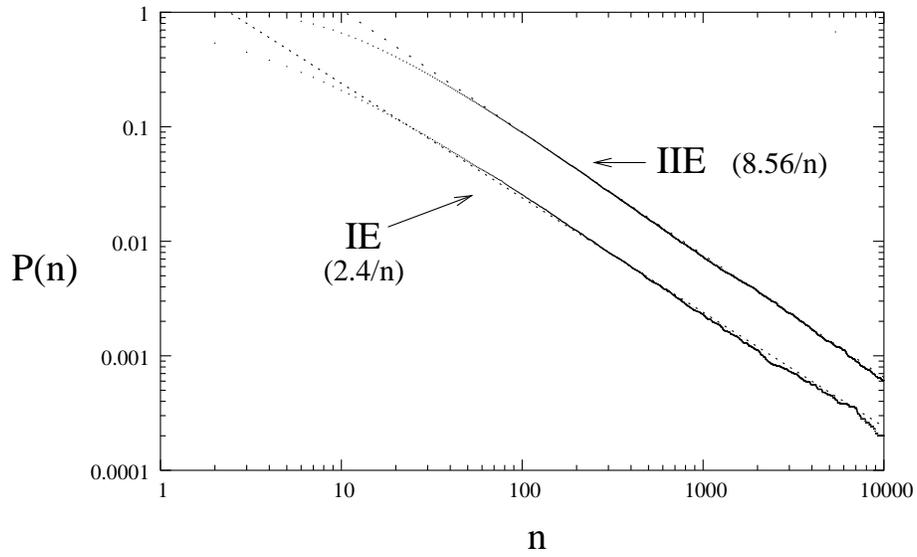,width=12cm}
\end{center}
\caption{Probability of getting a walk of size n on a square lattice
  for IE and IIE.}    \label{saw2d}
\end{figure}

IIE enhances the performance of the algorithm by increasing the
probability of connection between root and level $n$. For SAW on a
square lattice, Fig. \ref{saw2d} shows the probability of connection
$P(n)$ for IE and IIE both. $P(n)$ is roughly $3.5$ times bigger for
IIE. In two dimensions, $T_n$ is of order $0.12 n^2$ for IIE. In three
dimensions the  performance is even better and $T_n \approx 0.056
n^2$, which is roughly a factor of $7.5$ less than the time taken by
IE.

\begin{figure}
\begin{center}
\epsfig{figure=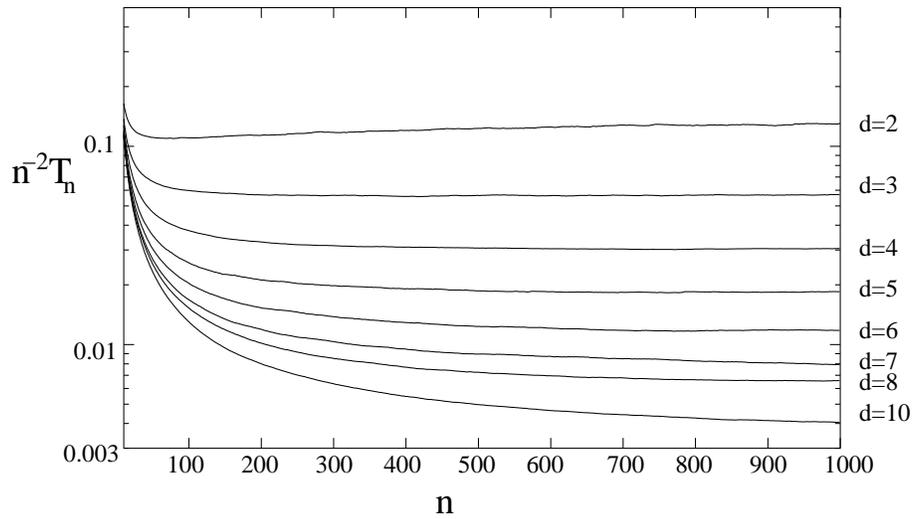,width=12cm}
\end{center}
\caption{$T_n/{n^2}$ of IIE Vs size n for SAW on a $2,3,4,5,6,7,8$ and
$10$ dimensional hyper cubic lattice.}    \label{sawen}
\end{figure}

In general we find on a d dimensional hypercube IIE has a efficiency
$T_n = a_d n^2$ where $a_d$ is a decreasing function of dimension  for
generating SAWs. Fig.\ref{sawen} shows the plot of IIE for dimensions
2 to 10. The memory requirement of the algorithm just increases
linearly with system size in all dimensions and we could perform
$10^5$ MC runs for walks up-to sizes $1000$ in few hours of computer
time on a Pentium-$4$ machine. We find that $a_d$ decreases as
$d^{-2}$  approximately, i.e the algorithm performs better with
increasing dimension.

We conclude that for IE and IIE for SAW, $T_n = a_d n^2$.  The
probability of connection between root and level $n$ does not depend
on $\gamma$. It depends only on the non-uniformity of the tree. The
genealogical tree is more uniform in higher dimensions  and the
constant  $a_d$ depends on dimension. For IE, the change in  $a_d$
with dimension is quite insignificant. But $a_d$ can be decreased
significantly by redistributing weights. This is a strong numerical
evidence that the performance  is always $O(n^2)$ independent of the
dimension and $\gamma$ for linear polymers.

A further enhancement can be achieved by choosing the pruning only
after looking deeper, but we found that because of the increase both
in the memory requirement and in the CPU time to generate one
configuration, there is no net gain over IIE.

\section{Lattice Animals and Branched Polymers}

In this section we will study the IE algorithm for branched
polymers. Since the efficiency of IE is polynomial in $n$ for linear
polymers, it seems plausible that it will be so also for branched
polymers. There are two important ways in which the genealogical tree
for branched polymers differ from that for linear polymers. There are
several equally reasonable, computationally easy to implement choices
of  rules to define parentage, and in all of them the degree of a node
is not bounded. The number of possible descendants of a node is of the
order of its perimeter sites and hence the maximum of the degree of
nodes  at level $n$ increases linearly with $n$. The average number of
descendants $\lambda$ is of $O(1)$, and the number of nodes with
large branching number is exponentially small. But this makes an
important  difference in the fluctuations of the number of animals of
a given size  generated in a given run.

The structure of genealogical tree for lattice animals is more complex
than for self-avoiding walks. We studied the algorithm on genealogical
tree  obtained by using Martin's labelling scheme \cite{martin}. We
have tried  two or three variations of the priority rules, and our
results are  insensitive to these changes.

\subsection{Lattice animals on a Binary tree}

We first discuss our results for the animals on a binary tree. This
simple case is more analytically tractable. The generating function of
total number of lattice animals on a binary tree is well known
\cite{grimmett} and it is $A(y) = \sum_0^{\infty} A_r y^r=
(1-\sqrt{1-4y})/2y$, where $A_r$ is the total number of animals with
$r$ sites. $A_r$ are the Catalan numbers, which come up in many  other
contexts in combinatorics \cite{stanley}. For large $r$ this gives
$A_r \sim 4^r r^{-\frac{3}{2}}$. The growth constant $\lambda$ in this
case is $4$.

\begin{figure}
\begin{center}
\epsfig{figure=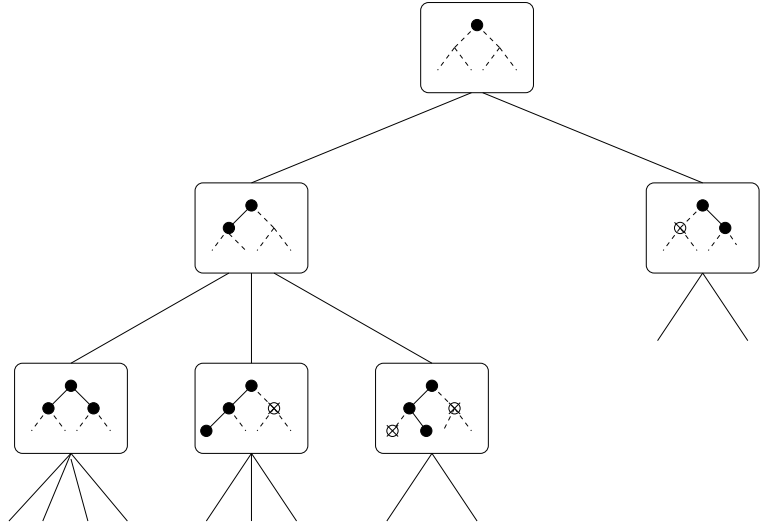,width=10cm}
\end{center}
\caption{ First few levels of the genealogical tree for lattice
animals  on a binary tree. Solid circles represent the occupied sites
and crossed  circles denote blocked sites on the Bethe lattice. Dotted
lines sketch the  underlying Bethe lattice, whereas solid lines
represent the bonds present. }
\label{rtree}
\end{figure}

The number of descendants of a node at level $r$ in the genealogical
tree for this problem lies between $2$ to $(r+1)$. In this case the
genealogical tree is easily characterised: The root site is a
$2-$node. A $k$-node has $k$ descendants, and the degree of these
descendants are $k+1,k,....3,2$ respectively. This is seen as follows:
the node corresponds to a branched polymer with $k$ unblocked
perimeter sites, which are ordered by some priority rule. The $m^{th}$
descendant of this node is a node of degree $(k+2-m)$ and corresponds
to first $(m-1)$ perimeter sites blocked, $m^{th}$ site occupied and
$(k-m)$ allowed for further occupation. Since on a binary tree every
site has two downward neighbours, hence we see that a $k$-node will
give rise to nodes with $k+1,k,.....2$ descendants. For example, in
Fig. \ref{rtree}, the top node corresponds to an animal of one site,
and has  two growth sites. If first of these two sites is occupied,
then the  corresponding animal has three growth sites. If it is
blocked it has  two growth sites and so on.

The total number of nodes at a level $r$ is equal to $A_r$. Let
$B_r(k)$ is the number of $k$-nodes at level $(r-1)$. We can determine
the distribution of the branching number. We find that $B_r(k)$
satisfy the following relation

\begin{equation}
B_r(k) = A_{r-2} - \sum_{s=2}^{k-2} B_{r-1}(s) \label{nodes}
\end{equation}

As $r \rightarrow \infty$, $1/4$ of the nodes at a level have $2$
offsprings and $1/4$ of the total nodes have $3$ offsprings. And level
$r$ has exactly one node with degree $(r+1)$. For $k \geq 4$, it can
be shown that in the asymptotic limit ($r \rightarrow \infty$), the
fraction of nodes having $k$ offsprings is $(k-1)/2^k$ for $r >> k$.

To find the efficiency factor $T_n$, we have to determine the
probability of connection of root to a level. If $P(k,r)$ is the
probability of a node with  $k$ offsprings to be connected to at-least
one node $r$ levels below it, then  $P(k,r)$ has a recursion

\begin{equation}
P(k,r+1) = 1- \prod_{s=2}^{k+1} (1-p P(s,r)) \mbox{ ,  k = 2 to }
\infty    \label{nrec}
\end{equation}

with initial conditions

\begin{equation}
P(k,1)= 1 \forall~~ k\geq 2                  \label{ic}
\end{equation}
and $p$ is the probability with which we choose any edge of the
tree. $P(2,r)$ will give the probability of connection of root to
level $r$ on the genealogical tree. Eq. (\ref{nrec}) is a nonlinear
equation. This equation can also be written as

\begin{equation}
1-P(k,r) = (1-P(k-1,r))(1-p P(k+1,r-1))~~~k>2            \label{nrec1}
\end{equation}

This equation is also valid for $k=2$ if we choose the convention that
$P(1,r) = p P(2,r-1)$.

These equations have the following properties:

\begin{enumerate}

\item For $p<1/4$, $P(k,r)$ tends to zero as $r$ tends to infinity
exponentially fast for any fixed $k$. In fact, if we consider $r$ as a
time like variable and $k$ as space like variable, then $P(k,r)$ has a
travelling front solution in this regime ($P(k,r) \cong F(k-vr)$).

\item For $p=1/4$, the velocity of travelling front goes to zero. The
distance moved by the front increases as $r^{1/3}$ and  $P(k,r) \sim
F(k-r^{1/3})$. As $F(x) \sim \mbox{exp}(x)$ for $x \rightarrow
-\infty$, this implies that $P(2,r) \sim \mbox{exp}(-cr^{1/3})$ for
large $x$.

\item For $p>1/4$, as $r$ goes to infinity, $P(k)$ tends to a non
trivial fixed point function $P^*(k)$ greater than zero.

\end{enumerate}

This may be seen as follows. The fixed point equation in terms of
fixed point variables $P^*(k)$ is

\begin{equation}
1-P^{*}(k) = (1-P^{*}(k-1))(1-p P^{*}(k+1))  \label{fp}
\end{equation}

Clearly, $P^{*}(k)=0$ $\forall k$ is a trivial fixed point of this
equation. For $p >1/4$, there is a non trivial fixed point with
$P^*(k)$ non zero monotonic increasing, with $ P^{*}(k) \approx
1-(1-p)^{k}$ for large $k$. However, a closed form solution for  any
$p>1/4$ is difficult.

On numerically iterating Eq. (\ref{nrec}) in $r$, we find that the
equation has a travelling front solutions for $p \leq 1/4$ and has
nontrivial fixed point for $p > 1/4$.

Eq (\ref{fp}) has two stationary solutions, i.e $P^{*}(k) =1 \forall
k$ and $P^{*}(k)=0 \forall k$. For $p \leq 1/4$, $P^{*}(k)=0$ is the
stable solution while $P^{*}(k)=1$ is an unstable solution. Our
initial conditions given by Eq (\ref{ic}) are steep. Starting with
these initial conditions, on numerical iteration we find that as $r$
increases, a front separating stable solution $P=0$ and  unstable
solution $P=1$ moves in the forward direction. From the translational
invariance of Eq. (\ref{nrec}) one expects a running wave solution. We
find that the front moves with a constant velocity and hence, $P(k,r)$
for large $k$ and $v$ must tend to the asymptotic form

\begin{equation}
P(k,r) \sim  F(k-vr)   \label{tf}
\end{equation}

We define $k^*(r)$, the width of the front by the equation,

\begin{equation}
P(k^*(r),r)=\frac{1}{2}
\end{equation}

\begin{figure}
\begin{center}
  \epsfig{figure=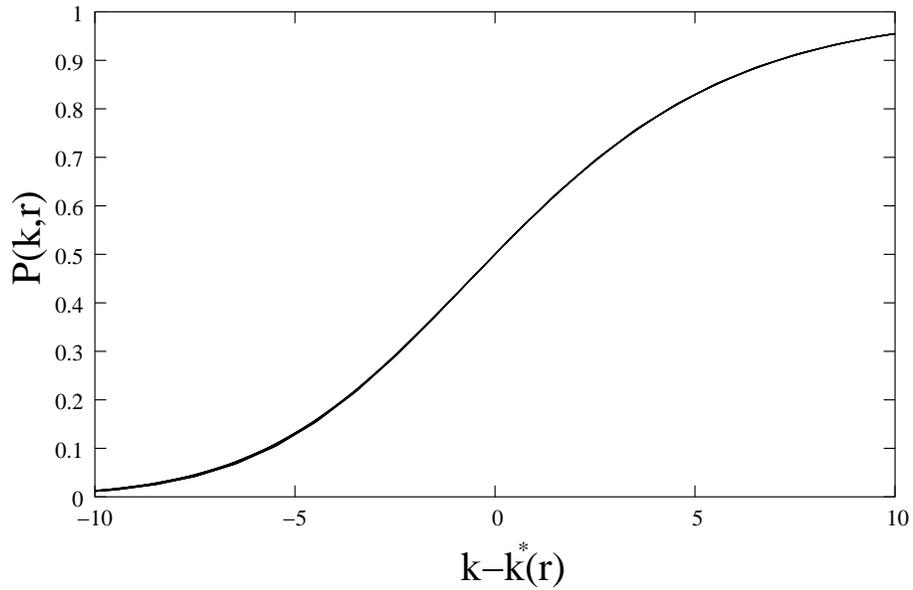,width=12cm}
\end{center}
\caption{Plot of $P(k,r)$ Vs scaled $k-k^{*}(r)$, for $p=0.25$ and
$p=0.25\pm 0.0001$ and $r=100,300$ and $600$. All the nine curves
collapse to the same front profile.}  \label{trfr}
\end{figure}

Fig. \ref{trfr} shows a plot of numerically determined $P(k,r)$ with
respect to $k-k^*(r)$ for $p$ near $1/4$. Curves for $p$ below, above
and at $p=1/4$ all collapse on the same line. Actually, a travelling
front for $P(k,r)$ as defined by Eq. (\ref{nrec1}) exists for all $k$,
$-\infty < k < \infty$, if we take boundary conditions such that
$P(-\infty,r)=0$ and $P(\infty,r)=1$.

\begin{figure}
\begin{center}
\epsfig{figure=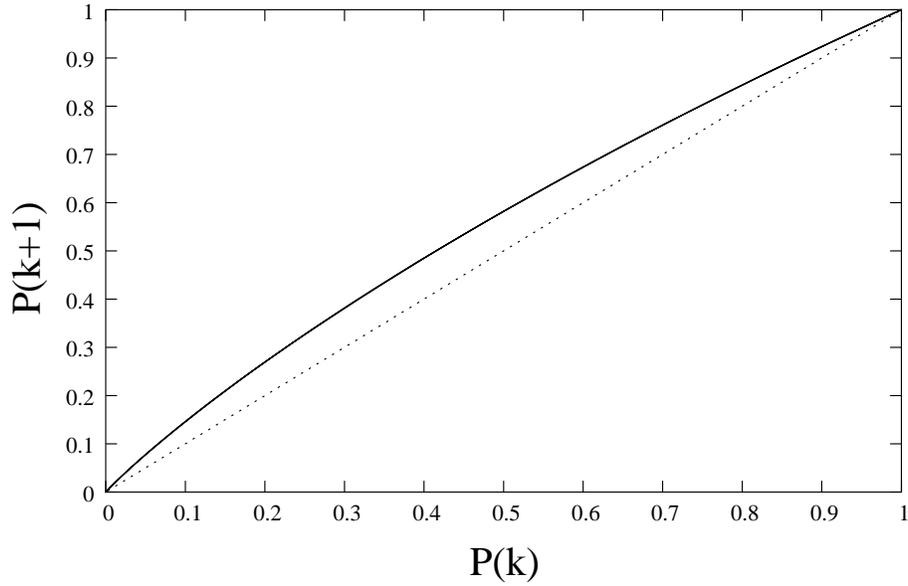,width=12cm}
\end{center}
\caption{ Plot of $P(k+1)$ as a function of $P(k)$ at $p=1/4$ for
$r=25000,26000,28000$ and $30000$. All the curves are very close and
approach a limiting form with increasing $r$. The dotted line is just
the line $x=y$.}  \label{pk1pk}
\end{figure}

At $p=1/4$, the velocity of the travelling front is zero. If we plot
$P(k+1,r)$ as a function of $P(k,r)$, we find that as $r$ increases
the graph approaches a limiting form. Thus for the asymptotic
wavefront,  $P(k+1,r)$ is a single valued nonlinear function of
$P(k,r)$. We have  plotted these values for different $r$ in
Fig. \ref{pk1pk} and  they all are very close and seem to lie on the
same curve. Hence if we start from a point on this curve and iterate
the fixed point equation Eq. (\ref{fp}) with $p=1/4$, we generate a
travelling front. We have not been able to deduce the functional  form
of this function, which corresponds to a first order difference
equation  for $P^{*}(k)$ from the second order equation
Eq. (\ref{fp}).  Eq. (\ref{fp}) turns out to be a stiff equation and
one has to be careful while iterating it in increasing $k$
direction. We iterated Eq. (\ref{fp}) starting with different sets of
values of $P^{*}(k+1)$ and $P^{*}(k)$ given by Fig. \ref{pk1pk} and
found the equation yields a travelling front same as the one shown in
Fig. \ref{trfr}.

We could not solve the full non-linear difference equation
Eq. (\ref{nrec}). Keeping only the terms linear in $P$ will give an
upper-bound on $P(k,r+1)$, i.e

\begin{equation}
P(k,r+1) \leq p \sum_{s=2}^{k+1} P(s,r) \label{lrec}
\end{equation}

We can represent this set of equations in matrix form also. Hence if
${\mathcal{P}}_r$ represents the infinite column array with $k^{th}$
entry being $P(k,r)$ then

\begin{equation}
{\mathcal{P}}_r \leq p^r M^r {\mathcal{P}}_0
\end{equation}
where $M$ is the transition matrix. If $\lambda_m$ is the largest
eigenvalue of $M$ then for $p < {1/{\lambda_m}}$, in the limit of $r
\rightarrow \infty$, ${\mathcal{P}}^{*}$ will be $0$, i.e $P^{*}(k) =
0$ for all $k$, and for $p < 1/{\lambda_m}$.

The elements $M_{i,j}$ of the transition matrix $M$ are such that,
$M_{i,j} =  1$ for $j \le (i+1)$ and $0$ otherwise. If we truncate $M$
beyond $n \times n$ ($M_n$), then the determinant $D_n$ of $M_n$ comes
out to be

\begin{equation}
D_n = A(\lambda) \left[\frac{1}{x_1^{n+1}} - \frac{1}{x_2^{n+1}}
\right]
\end{equation}
with $x_1,x_2 = \frac{-1}{2} \pm \sqrt{1-\frac{4}{\lambda}}$, and
$A(\lambda) = 1/\sqrt{1-\frac{4}{\lambda}}$ is a coefficient which
does not depend on $n$. Then equating $D_n =0 $ in the $ n \rightarrow
\infty$ limit gives $\lambda_m =4$. This implies that for $p < 1/4$,
$P(k,r)$ will decay exponentially with increasing $r$ and Eq
(\ref{lrec})  will work well. Hence, by definition percolation
threshold $p_c$ of  this tree is $1/4$.

The linearised recursion can be solved explicitly, and we get,

\begin{equation}
    P(k,r) = p^r \frac{\Gamma[k+2r-1]}{\Gamma[k+r-1] \Gamma[r+1]}
\end{equation}
		      which for large $r$ gives
\begin{equation}
P(k,r) \sim \frac{1}{4 \sqrt{\pi r}} \mbox{exp}\left[\ln 2 \left(k+r
			  \frac{\ln(4p)}{\ln(2)}\right)\right]
			  \label{sl}
\end{equation}

If we assume a travelling front solution of kind $P(k,r) \propto
\mbox{exp}(\lambda(k-vr))$ to be valid in the tail of the
distribution, then  substituting in linearised recursion
(Eq.(\ref{lrec})), for a given $p$ we get a spectrum of travelling
wave like solutions parametrised by $\lambda$ with the velocity $v$ of
the front given by

\begin{equation}
v = \frac{1}{\lambda} \mbox{ln} \frac{1-\mbox{exp}(-\lambda)}{p}-1
\label{vel}
\end{equation}
In this case, it is known that the front actually chooses a unique
velocity given by minimum of right hand side of Eq. (\ref{vel}) with
respect to $\lambda$ \cite{saarloos}. The front velocity is given by

\begin{equation}
v^* = \frac{2\mbox{exp}(-\lambda^{*})-1}{1-\mbox{exp}(-\lambda^{*})}
\label{vel1}
\end{equation}

where $\lambda^{*}$ is the solution of the transcendental equation

\begin{equation}
\frac{-1}{\lambda^{*}} \mbox{ln} \frac{1-\mbox{exp}(-\lambda^{*})}{p}
+ \frac{\mbox{exp}(-\lambda^{*})}{1-\mbox{exp}(-\lambda^{*})} = 0
\end{equation}

Near $p=1/4$, we can take $v \approx \mbox{ln}(4p)/\mbox{ln}(2)$ and
$\lambda \approx \mbox{ln}2$. Travelling front solutions have been
found in a large variety of problems in physics \cite{tf}.

 The linearisation of Eq (\ref{nrec}) would be valid only for $p \leq
  1/4$ and $k < k_o(r)$. Beyond that, linear solution will grow beyond
  one whereas the solution of the full nonlinear equation will
  saturate to $1$. Here $k_o(r)$ is the value of $k$ at which $P(k,r)$
  given by Eq. (\ref{sl}) becomes of $O(1)$ and is equal to

\begin{equation}
k_o(r) = \frac{-r \mbox{ln}(4p)}{\mbox{ln}2}~~~,~ for ~~~ p<\frac{1}{4}
\end{equation}

   At $p=1/4$, the asymptotic velocity of the front is zero and  the
front advances as a sub linear power of $r$. This is the  critical
point of the percolation on this tree, and Eq. (\ref{sl}) gives a
algebraic decaying solution for sufficiently small $k$. This is only
an upper bound to the actual value. On numerically iterating
Eq. (\ref{nrec}) for  $r$ upto order $10^4$, we found unexpectedly
that it decays as a stretched exponential in $r$.

The fixed point equation as given by Eq. (\ref{fp}) is again a
nonlinear equation. To find the dependence of probability of
connection of root, $P(2,r)$, on the width of the front we solved the
linearised fixed point equation. On solving, we find that it goes as
$2^{-k^*(r)}$ for large $r$, where $k^*(r)$ is the width of the
distribution. Hence in general, $P(2,r) \sim \mbox{exp}(-a k^*(r))$.

\begin{figure}
\begin{center}
  \epsfig{figure=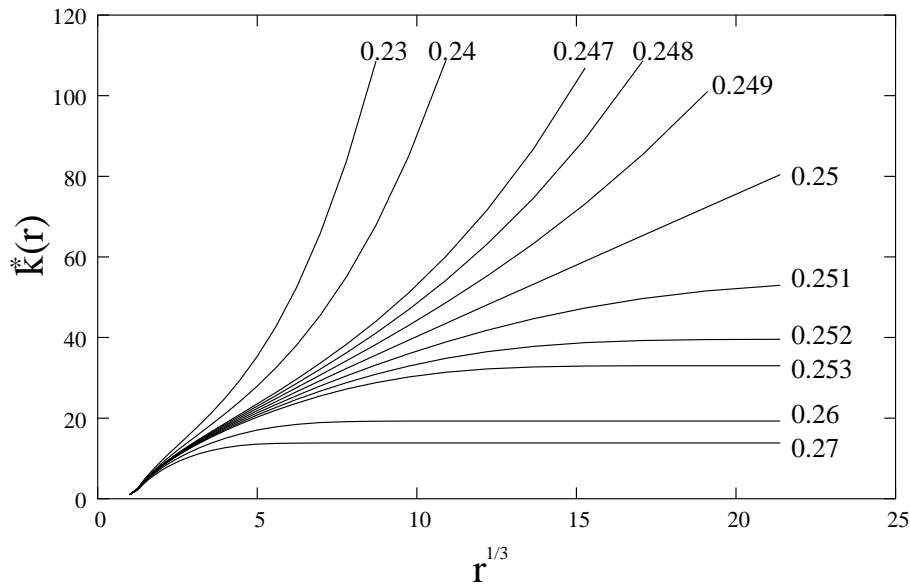,width=12cm}
\end{center}
\caption{The width $k^*(r)$ of the travelling front as a function of
$r^{1/3}$ for different values of $p$. The value of $p$ increases from
left to right. Curves of left of $p=1/4$ are for $p<1/4$ and the ones
on right are for $p>1/4$. For $p=1/4$ the graph approaches a straight
line as $r \rightarrow \infty$.}
\label{kr}
\end{figure}

\begin{figure}
\begin{center}
  \epsfig{figure=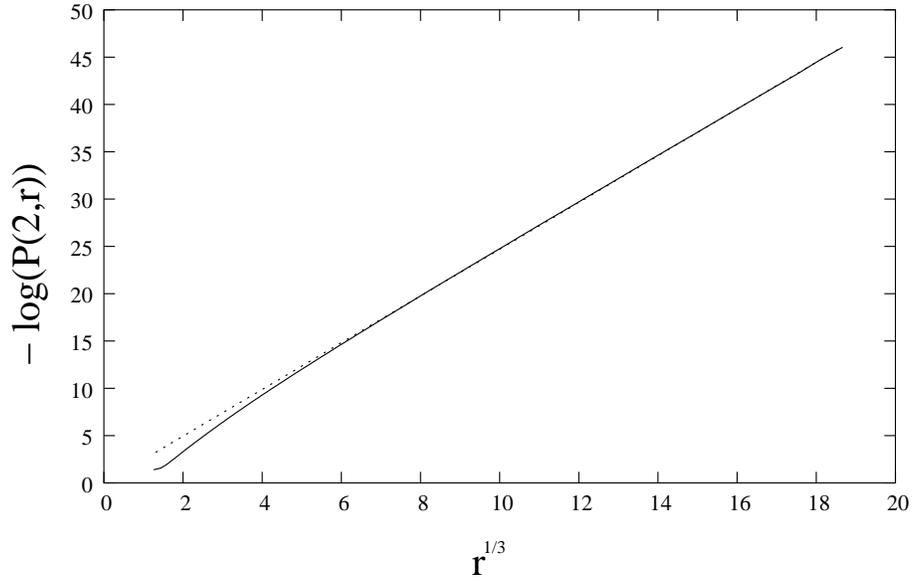,width=12cm}
\end{center}
\caption{Thick line is the plot of $-\mbox{log}(P(2,r))$ as a function
of $r^{1/3}$, when $p$ is taken to be $1/4$. The dotted line is a
straight line of slope $2.47$.}  \label{P2rt}

\end{figure}

We further studied the width $k^*(r)$ of the front as a function of
$r$ for different values of $p$. At $p=1/4$ we found $k^*(r) \sim
r^{\frac{1}{3}}$. Fig. \ref{kr} shows a plot of $k^{*}(r)$ as a
function of $r^{1/3}$. For $p=1/4$, the plot is a straight line. This
implies that $P(2,r) \sim \mbox{exp}(-c r^{1/3})$ at $p=1/4$. For
$p<1/4$, $k^{*}(r)$ varies linearly with $r$ and tends to a constant
for $p>1/4$. We can directly iterate Eq. (\ref{nrec}). In Fig
\ref{P2rt} we have plotted $-log(P(2,r))$ as a function of  $r^{1/3}$
which comes out to be a straight line. Fig \ref{kr} and
Fig. \ref{P2rt} are strong numerical evidence that the probability of
connection goes as $\mbox{exp}(-c r^{\alpha})$ for branched polymers
on binary tree. Our numerical studies give $\alpha = 0.333 \pm 0.005$
and $c = 2.47 \pm 0.01$.

\subsection{Heuristic argument for the stretched exponential 
behaviour of $P(n)$ at $p=1/4$}

We now present a heuristic argument to understand why $k^{*}$ varies
as $r^{1/3}$ at $p_c$. Let us consider a genealogical tree of lattice
animals on a binary tree, in which nodes with more than $k$
descendants are deleted. We denote the probability that the maximum
degree of a node connected to root  down to level $r$ is $k_m$, by
$H_2(k_m)$ and the probability that a $k_m$ node is connected to at
least one node $r$ level down on the truncated genealogical tree by
$J_{k_m}(r)$.

Now on a truncated tree, transition matrix $M$ is no longer
infinite. It is now a $k_m \times k_m$ matrix with $M_{i,j}=1$ for $j
\leq (i+1)$ and $0$ otherwise. Here $M_{i,j}$ represents the $i^{th}$
row and $j^{th}$ column entry of $M$, and we find the critical value
of $p$ which is just inverse of the largest eigenvalue of $M$ to be a
function of $k_m$ and is equal to

\begin{equation}
p_c(k_m) = \frac{1}{4}
\left(1+{tan^2}\left(\frac{{\pi}}{k_m+1}\right)\right)
\end{equation}

For $p<p_c(k_m)$, $J_{k_m}(r)$ decays exponentially with $r$. In large
$r$ limit it is given by

\begin{equation}
J_{k_m}(r) \sim \mbox{exp}(r log(p/p_c(k_m)))
\end{equation}

At $p=1/4$, we get $J_{k_m}(r) \sim \mbox{exp}(-br/k_m^2)$, where $b$
is a constant.

It is easy to get a lower bound on $H_2(k_m)$, as a order $k_m$ node
 occurs first time at level $k_m$ and probability of connection of
 root to this node is $p^{k_m}$. Hence

\begin{equation}
H_2(k_m) \geq p^{k_m} = 4^{-k_m} \label{upb}
\end{equation}

Hence, since $p=1/4$ is less than $p_c(k_m)$ for any finite $k_m$,
$J_{k_m}(r)\sim  \mbox{exp}(-br/k_m^2)$, where $b = \pi^2$.  Since
$H_2(k_m) \ge\mbox{exp}(-a_p k_m)$, for large $r$ we get

\begin{equation}
P(2,r) \geq \max_{k_m} \left[\mbox{exp}\left(-a_p
k_m-\frac{br}{k_m^2}\right)\right]
\end{equation}

which gives

\begin{equation}
P[2,r] \geq exp(-c r^{1/3})
\end{equation}

where $c= \frac{3}{2} (2ba_p^2)^{\frac{1}{3}}$. If we take $H_2(k_m)$
to be as given by Eq. (\ref{upb}), we get an lower bound on
$P(2,r)$. Taking $b=\pi^2$ and $a_p=\mbox{log}4$ we get $c =
\frac{3}{2} (2 \pi^2 {log^2{4}})^{1/3}= 5.04$. This should be compared
with the numerical estimate $c \cong 2.47$.

Thus our numerical simulations and qualitative arguments show that
probability of connection goes down as a stretched exponential at
$p=1/4$, the $p_c$ of the genealogical tree of lattice animals on
binary tree as opposed to $r^{-1}$ decay for linear polymers. So if we
chose $p_i =1/4$ $\forall i$, then $\langle{X_r}\rangle \sim r^{-3/2}$
and hence the average computer time to generate one statistically
independent sample of size $r$, $T_r$ would go as $\mbox{exp} (c
r^{1/3})$ to leading order.

Clearly the algorithm is not working well and one would like to
enhance its efficiency if possible. We tried to study the algorithm by
choosing $p_i$ such that its asymptotic value is $1/4$. We chose $p_i
= \frac{1}{4} (1+\frac{x}{i^{m}})$ and studied $T_r$ as a function of
$x$ and $m$.

As argued earlier, taking $m=1$, we can change $\tau_r$ and $P(r)$ by
multiplicative factors which are powers of $r$. This will not make
much of a difference, as the leading dependence remains
$\mbox{exp}(cr^{1/3})$. Using $m<1$, seems to be more interesting.

\begin{figure}
\begin{center}
  \epsfig{figure=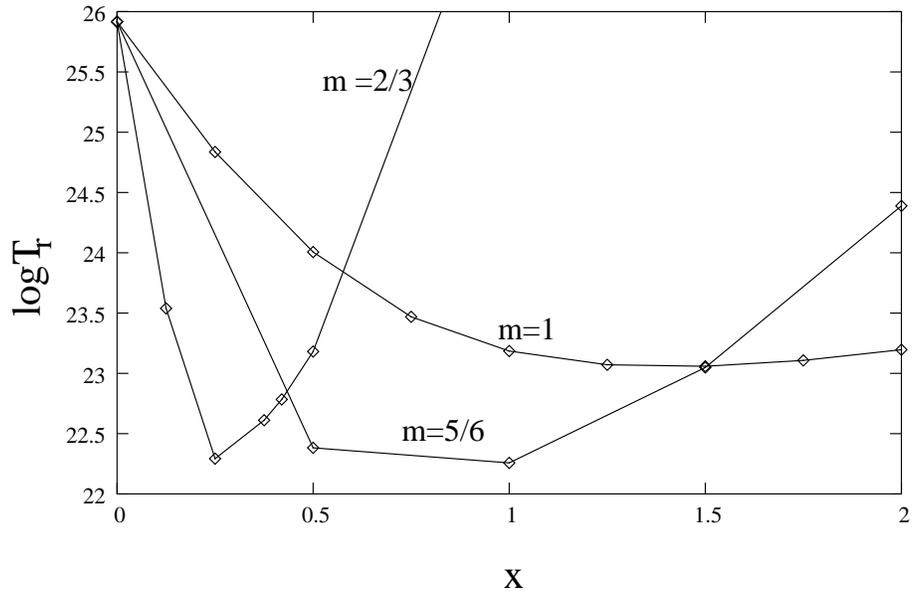,width=12cm}
\end{center}
\caption{Plot of $\mbox{log}T_n$ for $m=2/3,5/6$ and $1$ as a function
of $x$ for $n=1000$.}       \label{plax}
\end{figure}

For $m<1$, the average CPU time per Monte-Carlo run would vary as
 $exp(x r^{1-m})$.  In case of linear polymers, we saw that time
 complexity of the algorithm for $m=1$ for any $x$ is polynomial in
 $r$. Hence, $m<1$ was clearly a bad choice. But in the case of
 lattice animals, this increase in numerator is exactly cancelled  by
 a corresponding increase in $P(r)$. For $2/3 \leq m < 1$, $\tau_r$
 increases as $\mbox{exp}(x r^{1-m})$ and $P(r)$ varies as
 $\mbox{exp}(-c r^{1/3}+xr^{1-m})$ to leading order for large
 $r$. These cancel to give $T_r \sim \mbox{exp}(cr^{1/3})$ independent
 of $m$. To monitor the behaviour of various prefactors,  we study
 this numerically. Fig. \ref{plax} shows plot of $T_r$ for $r=1000$,
 for $m=2/3,5/6$ and $1$ as a function of $x$. For $1 \ge a \ge 2/3$,
 to leading order $T_r$ goes as $\mbox{exp}(cr^{1/3})$, but there
 exist a non trivial value of $x$ at which $T_r$ is minimum for a
 given $m$. If we look at $T_r$ at best value of $x$ for $m=2/3,5/6$
 and $1$, we find that as $r$ increases the difference is not
 significant.

Hence we conclude that to leading order, $T_r \sim
\mbox{exp}(cr^{1/3})$, for the best choice of $p$. For all $2/3 \leq m
\leq 1$, there exist a range of $x$ for which the time complexity of
the algorithm will remain qualitatively the same.

\subsection{Lattice Animals on a 2 dimensional square lattice}

We also studied the efficiency for lattice animals on a square
lattice. From exact series enumeration the $A_r$ is known to vary as
$\lambda^r$ with $\lambda \approx4.06257$ \cite{jensen}. In this case
also the number of offsprings a node at level $r$ can have is $O(r)$
and the genealogical tree in this case though more complicated, is
qualitatively similar. Numerically, we find that the probability
distribution of number of descendants $k$ (of a randomly chosen node)
has a maximum at $k=4$, with $Prob(k=4) \approx 1/4$. We enumerated
lattice animals up-to sizes 1000 using IE with $10^6$ Monte-Carlo
runs. It took time of order one day on a Pentium-4 machine. With IIE
we generated samples of size 2000 with $2 \times 10^6$ Monte-Carlo
runs in 2-3 days time. These sizes are of same order as those produced
using the cut and paste type algorithms.

\begin{figure} 
\begin{center}
  \epsfig{figure=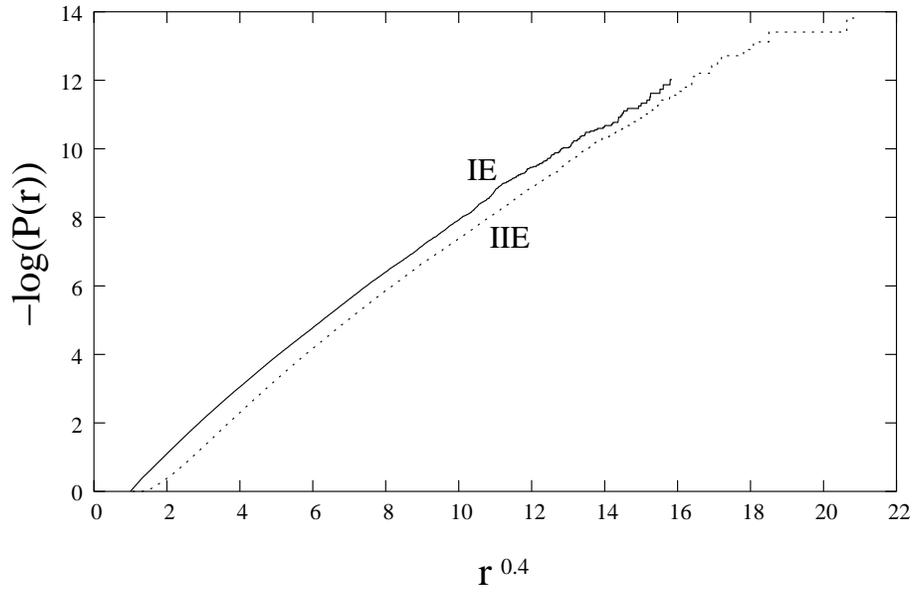,width=12cm}
\end{center}
\caption{Plot of $-log(P(2,r))$ Vs $r^{0.4}$ for lattice animals on a
square lattice with IE and IIE}  \label{la2d}

\end{figure}

\begin{figure} 
\begin{center}
  \epsfig{figure=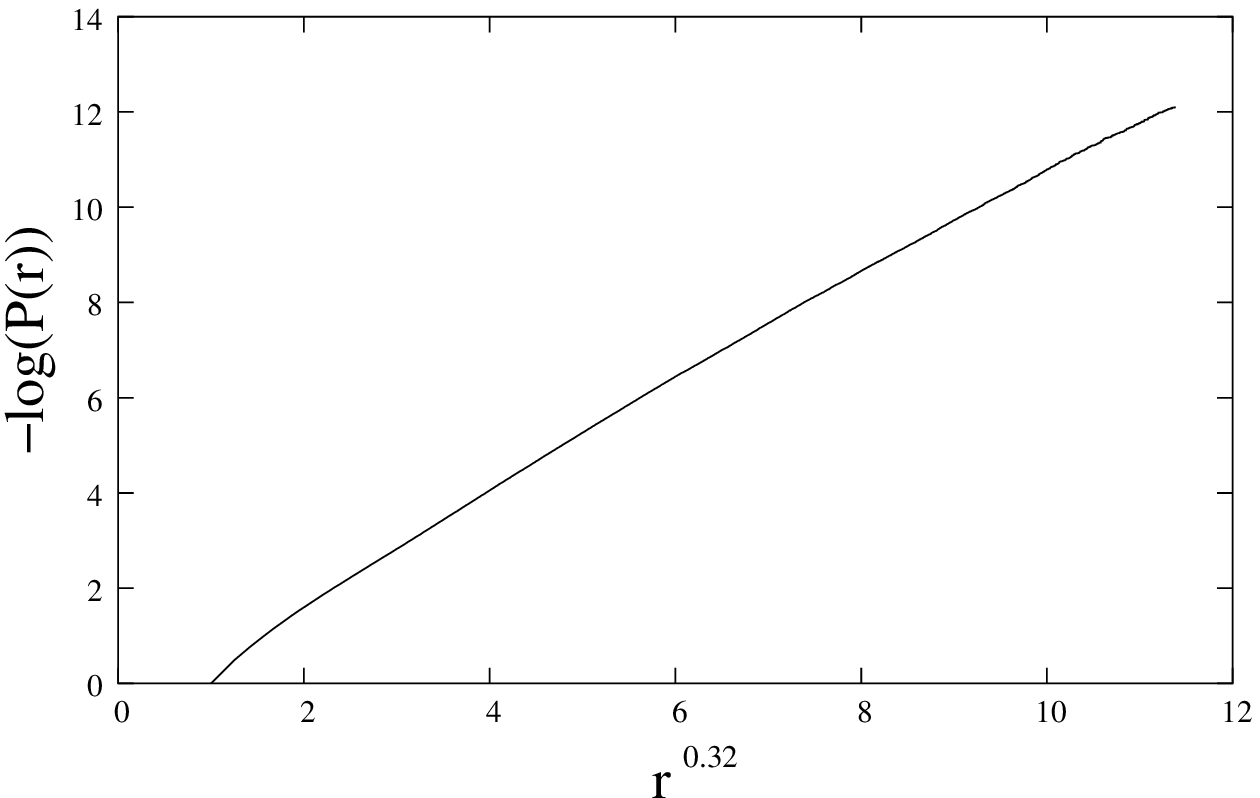,width=12cm}
\end{center}
\caption{Plot of $-\mbox{log}(P(r))$ Vs $r^{0.32}$ for directed
animals with IIE.}      \label{da2d}

\end{figure}

In this case, we find that $P(r)$ has the stretched exponential form
$P(n) \sim \mbox{exp}(-c n^{\alpha})$, with $\alpha \approx 0.4$ for
both  IE and IIE. Fig. \ref{la2d} shows $[-\mbox{log}P(r)]$ varies
approximately  linearly with $r^{0.4}$. We also studied the directed
lattice animals (DA)  on a square lattice. In this case we find that,
$\alpha=0.32\pm 0.02$  (Fig. \ref{da2d}).

\section{Discussion}

We find the efficiency of IE to be different for linear and branched
polymers. This is due to the fact that genealogical tree for the
latter is much more non uniform.

For self avoiding walks, in any dimension, the time to generate an
independent sample of $n$ steps $T_n \sim a_d n^2$, independent of
dimension for both IE and IIE. For IE there is no significant change
in $a_d$ with dimension. But for IIE $a_d \sim d^{-2}$. In the
limiting case of SAW on binary tree $T_n=n$ for IIE.

For branched polymers $T_n$ increases as $\mbox{exp}(c n^{\alpha})$
with $0<\alpha<1$ in all dimensions for both IE and
IIE. Redistributing  weight does not change the value of $\alpha$. IIE
works better than IE,  but the difference is only in the coefficient
$c$. The exponent $\alpha$  depends weakly on the dimension, its
relation to the usually studied exponents of  the branched polymer
problem eg. $\theta$, $\nu$ is not clear at present.

As discussed earlier, the genealogical tree for cluster enumeration is
not unique and one might argue that Martin's scheme is not the optimal
choice. We tried to generate the genealogical tree using some
variations of this rule, but we did not find any significant change in
efficiency of the algorithm.

For branched polymers, the degree of a node in the genealogical tree
is not bounded, and the maximum degree increases with depth of the
genealogical tree. However, the fractional number of nodes with  high
degree is very small. For genealogical tree corresponding to animals
on a binary tree we find the fractional number of $k$-nodes goes down
exponentially with $k$ for large $k$(Eq. (\ref{nodes})). Similar,
behaviour was observed for branched polymers and directed branched
polymers on a square lattice numerically. It is surprising  that even
an exponentially rare distribution of nodes with large degree  seems
to be enough to change the behaviour of efficiency of the algorithm on
the tree.

In the case of branched polymers, we found that the $T_n$ for IE
varies  as $\mbox{exp}(cn^{\alpha})$ with $0<\alpha <1$. While this is
not very good, one can find problems for which IE's performance is
even worse with  $\alpha=1$. As an example, consider self avoiding
walks on a disordered  lattice, obtained by removing a fraction
$(1-u)$ of bonds at random from  a square lattice. It is known that
the average number of self avoiding  walks of length $n$ varies as $(u
\lambda)^n$ \cite{barat},  where $\lambda$ is the growth constant of
the self avoiding walks  on the same lattice with $u=1$. Hence the
growth constant of the  corresponding genealogical tree would also be
$u \lambda$. Now if we  consider a square lattice, the $\lambda
\approx 2.638$ and the bond  percolation threshold is $1/2$. For
$1/\lambda < u<1/2$,  all clusters would be finite with probability 1,
and the probability  that cluster contains $n$ sites would decrease
exponentially with $n$.  In this case, IE will be inefficient and even
for best choice,  $T_n$ will vary as $\mbox{exp}(cn)$.

One could argue that IE is a rather inefficient algorithm, which gives
reasonable performance only for a small selected set of problems.  We
do not think so. In fact, the causes that make IE inefficient are
also operative in the much larger class of genetic type
algorithms. The  high degree of correlations between different samples
generated is a common  feature of many of these algorithms which
employ pruning and enrichment.  For example, one could expect a
similar behaviour to occur in the  Berreti-Sokal algorithm
\cite{berreti}, for branched polymers. The  correlations arise because
in all such `evolutionary' type algorithms  different samples
generated often share a common ancestor in the past.  Whether our
results can be generalised to a larger class of PERM type  algorithms
seems to be an interesting question for further study.

{\bf{Acknowledgements}}

We thank M.Barma for a careful reading of the manuscript, and the
referee  for many helpful remarks, which have helped improve our
presentation.

\section{Appendix}

As discussed in Sec.1, to enumerate all allowed configurations on a
computer, one need a good exact enumeration algorithm which would
generate all possible configurations exactly once, without needing to
refer to what has been generated previously. Hence, one has to label
the $n$-point configurations such that for any $n$-point configuration
the labelling is unique and on removing the last added site we must
get an allowed $(n-1)$-point configuration. For the self avoiding
walks this can be easily achieved by labelling the first point of walk
as 1, the second 2, and so on. But usually such natural choice of
labelling doesn't exist for most problems. For lattice trees and
animals, Martin discusses this in detail \cite{martin}. Here we
describe briefly his algorithm for labelling a $n$-cluster.

\begin{itemize}

\item Choose a rule for ordering the neighbours of any given site.
For example, for DA on a square lattice (Fig.1 ), we chose the rule
that the upward neighbour is labelled before the right neighbour.  For
lattice animals on the binary tree we choose left neighbour  before
the right neighbour(Fig. 6).

\item  We label the root as $1$ and its neighbours are labelled
$2,3,4....$  in the order according to the priority rule.

\item When all points adjacent to point $1$ have been labelled, label
any still unlabelled points adjacent to point $2$ according to the
priority  rule and then of point $3$ and so on. This labeling hence
induces a tree  structure on the cluster which is the genealogical
tree.

\end{itemize}

The labelling described above is just one way of labelling the
configurations. One can invent many other labelling schemes, which
would give rise to different genealogical tree. But we find that the
nature of genealogical tree depends on the underlying problem and not
on the rules of labelling.

\section*{Figure Captions}

\begin{itemize}
\item Figure1:An example of a genealogical tree. The numbers labelling the
sites indicate the order in which they are added ($1$ represents the
root site). The tree shown is for directed lattice site animals on a
square lattice.

\item Figure2:Plot of optimum values of $p_i$ on a $(23,233)$ tree of depth
$30$

\item Figure3: $T_n/n^2$ of IE as a function of size $n$ for SAW on a $2$,
$3$ and $4$ dimensional hyper cubic lattice. The lower most graph 
is for SAW on a square lattice and middle one in $3-d$ and the 
topmost is for $4-d$.

\item Figure4:Probability of getting a walk of size n on a square lattice
  for IE and IIE.

\item Figure5:$T_n/{n^2}$ of IIE Vs size n for SAW on a $2,3,4,5,6,7,8$ and
$10$ dimensional hyper cubic lattice.

\item Figure6:First few levels of the genealogical tree for lattice
animals  on a binary tree. Solid circles represent the occupied sites
and crossed  circles denote blocked sites on the Bethe lattice. Dotted
lines sketch the  underlying Bethe lattice, whereas solid lines
represent the bonds present.

\item Figure7:Plot of $P(k,r)$ Vs scaled $k-k^{*}(r)$, for $p=0.25$ and
$p=0.25\pm 0.0001$ and $r=100,300$ and $600$. All the nine curves
collapse to the same front profile.

\item Figure8:Plot of $P(k+1)$ as a function of $P(k)$ at $p=1/4$ for
$r=25000,26000,28000$ and $30000$. All the curves are very close and
approach a limiting form with increasing $r$. The dotted line is just
the line $x=y$.

\item Figure9:The width $k^*(r)$ of the travelling front as a function of
$r^{1/3}$ for different values of $p$. The value of $p$ increases from
left to right. Curves of left of $p=1/4$ are for $p<1/4$ and the ones
on right are for $p>1/4$. For $p=1/4$ the graph approaches a straight
line as $r \rightarrow \infty$.

\item Figure10:Thick line is the plot of $-\mbox{log}(P(2,r))$ as a function
of $r^{1/3}$, when $p$ is taken to be $1/4$. The dotted line is a
straight line of slope $2.47$.

\item Figure11:Plot of $\mbox{log}T_n$ for $m=2/3,5/6$ and $1$ as a function
of $x$ for $n=1000$.

\item Figure12:Plot of $-log(P(2,r))$ Vs $r^{0.4}$ for lattice animals on a
square lattice with IE and IIE.

\item Figure13:Plot of $-\mbox{log}(P(r))$ Vs $r^{0.32}$ for directed
animals with IIE.

\end{itemize}

\begin{thebibliography}{999}

\bibitem{sokal} A D Sokal, Monte-Carlo Methods for the Self Avoiding
Walk 1995, in Monte Carlo and Molecular Dynamics Simulations in
Polymer  Science, ed. K. Binder, Oxford University Press New York
47-124,  hep-lat/9405016

\bibitem{rouse} M Doi and S F Edwards 1986, The Theroy of Polymer
Dynamics,  Clarendon Press-Oxford.

\bibitem{reptation} F T Wall and F Mandel 1975, Macromolecular
dimensions  obtained by an efficient Monte Carlo method without sample
attrition  J. Chem. Phys. {\bf 63} 4592-4595.

\bibitem{pivot} N Madras and A D Sokal 1988, The Pivot Algorithm: A
Highly Efficient Monte Carlo Method for the Self-Avoiding Walk,
J. Stat. Phys. {\bf 50} 109

\bibitem{cutpaste} S Caracciolo, A. Pelissetto and A D Sokal 1990, A
 Nonlocal Monte Carlo Algorithm for Self-Avoiding Walks with Fixed
 Endpoints, J. Stat. Phys. {\bf 60} 1

\bibitem{wall} F T Wall and J J Erpenbeck 1959, New method for the
statistical computation of polymer dimensions, J. Chem. Phys. {\bf 30}
634-637.

\bibitem{grassberger} P Grassberger and W Nadler 2000, Go with  the
winners - Simulations, cond-mat/0010265, Proceedings der
Heraeus-Ferienschule "`Vom Billiardtisch bis Monte Carlo: Spielfelder
der statistischen Physik"', Chemnitz, October 2000;

\bibitem{saw} E J Janse van Rensburg, S.G.Whittington, and N. Madras
1990, The pivot algorithm and polygons: results on the FCC lattice,
J. Phys. A,{\bf 23} 1589.

\bibitem{kennedy} T Kennedy 2002, A faster implementation of the pivot
algorithm for self-avoiding walks, J. Stat. Phys. {\bf 106} 407-429

\bibitem{berreti} A Berreti and A.D.Sokal 1985, New Monte Carlo method
for Self-Avoiding Walk, J. Stat. Phys. {\bf 40} 483

\bibitem{rensburg} E J Janse van Rensburg and N Madras, A nonlocal
Monte Carlo algorithm for lattice trees, J. Phys. A:Math. Gen.  {\bf
25} 303-333(1992); Metropolis Monte Carlo simulation of lattice
animals,  J Phys. A:Math. Gen. {\bf 30} 8035-8066 (1997);  E J Janse
van Rensburg and A Rechnitzer 2003, High precision canonical  Monte
Carlo determination of the growth constant of square lattice  trees,
Phys. Rev E {\bf 67} 0361161-0361169

\bibitem{you} S You and E J Janse van Rensburg 2001, Adsorbing trees
in two dimensions: A Monte Carlo study, Phys. Rev. E vol. {\bf 64}
0461011-0461019.

\bibitem{grassberger1} H P Hsu, W Nadler and P Grassberger 2004,
Simulations of lattice animals and trees, cond-mat/0408061

\bibitem{redner} S Redner and P.J. Reynolds 1981, Position-space
  renormalisation group for isolated polymer chains, J. Phys. A {\bf
  14} 2679.

\bibitem{dhar} D Dhar and P M Lam 1986, A Monte Carlo method for
series expansions, J. Phys. A: Math. Gen. {\bf 19} L1057-1061

\bibitem{lam} P M Lam 1986, Monte Carlo study of lattice animals in
$d$ dimensions, Phys. Rev. A, Vol. {\bf 34} 2339-2345.

\bibitem{martin} J. L. Martin 1974, Computer Techniques for Evaluating
Lattice Constants, Phase Transitions and Critical Phenomena ,Eds. Domb
and Green, vol. {\bf 3}.

\bibitem{harris} T E Harris 1963, Theory of branching processes,
Springer-Verlag Berlin.

\bibitem{grimmett} G Grimmett 1989, Percolation, Springer-Verlag

\bibitem{madras} N Madras and G Slade 1993, The Self Avoiding Walk,
Birkhauser Boston.

\bibitem{nienhuis} B. Nienhuis 1982,  Exact Critical Point and
Critical Exponents of O(n) Models in Two Dimensions,
Phys. Rev. Letts. {\bf 49}, 1062

\bibitem{jensen1} I Jensen and A J Guttmann 1999, Self-avoiding
polygons on the square lattice, cond-mat/9905291

\bibitem{stanley} R P Stanley 1999, Enumerative Combinatorics, Vol.2
Chapter 6, Cambridge University Press, Cambridge-New York.

\bibitem{saarloos} Wim van Saarloos 2003, Front propagation into
unstable states, Phys. Rep. {\bf 386} 29

\bibitem{tf} E Brunet and B Derrida 1997, Shift in the velocity of a
 front due to a cutoff, Phys. Rev. E {\bf 56} 2597; S N Majumdar and
 P.L Kaprivsky 2003, Extreme Value Statistics and Travelling Fronts:
 Various Applications, Physica A {\bf 318} 161

\bibitem{jensen} I Jensen 2001, Enumerations of lattice animals and
trees, J. Stat. Phys. Vol. {\bf 102} 865-881.

\bibitem{barat} K. Barat and B. K. Chakrabarti 1995, Statistics of
self-avoiding walks on random lattices, Phys. Rep. {\bf 258} 377.

\end{thebibliography}
\end{document}